\documentclass[aps,twocolumn,superscriptaddress,showpacs,showkeys,amsmath,amssymb,floatfix]{revtex4}%preprint <-> twocolumn
\usepackage{booktabs,graphicx,mathrsfs,verbatim}
\usepackage[colorlinks=true,citecolor=blue,filecolor=blue,linkcolor=blue,urlcolor=blue,pdftex]{hyperref}
\usepackage[usenames,dvipsnames]{color}
\usepackage{ulem}

\newcommand{\1}[1]{\, \mathrm{#1}} % unit(y ;-)
\newcommand{\n}[1]{\mathrm{#1}} % normal (roman) text in math mode
\newcommand{\dd}{\, \mathrm{d}}

\newcommand{\percent}{\%}

\newcommand{\arxiv}[1]{\href{http://arxiv.org/abs/#1}{\texttt{arXiv:#1}}}

\newcommand{\ambe}{${}^{241}$AmBe}

\newcommand{\assergi}{\affiliation{INFN Laboratori Nazionali del Gran Sasso, Assergi, 67100, Italy}}

\newcommand{\columbia}{\affiliation{Physics Department, Columbia University, New York, NY 10027, USA}}
\newcommand{\coimbra}{\affiliation{Department of Physics, University of Coimbra, R. Larga, 3004-516, Coimbra, Portugal}}
\newcommand{\heidelberg}{\affiliation{Max-Planck-Institut f\"ur Kernphysik, Saupfercheckweg 1, 69117 Heidelberg, Germany}}
\newcommand{\houston}{\affiliation{Department of Physics, Rice University, Houston, TX 77005 - 1892, USA}}

\newcommand{\losangeles}{\affiliation{Physics \& Astronomy Department, University of California, Los Angeles, USA}}
\newcommand{\mainz}{\affiliation{Institut f\"ur Physik, Johannes Gutenberg Universit\"at Mainz, 55099 Mainz, Germany}}
\newcommand{\munster}{\affiliation{Institut f\"ur Kernphysik, Wilhelms-Universit\"at M\"unster, 48149 M\"unster, Germany}}
\newcommand{\shanghai}{\affiliation{Department of Physics, Shanghai Jiao Tong University, Shanghai, 200240, China}}
\newcommand{\subatech}{\affiliation{SUBATECH, Universit\'e de Nantes, 44307 Nantes, France}}
\newcommand{\weizmann}{\affiliation{Department of Particle Physics and Astrophysics, Weizmann Institute of Science, 76100 Rehovot, Israel}}
\newcommand{\zurich}{\affiliation{Physics Institute, University of Z\"{u}rich, Winterthurerstr. 190, CH-8057, Switzerland}}

\begin{document}

\title{Likelihood Approach to the First Dark Matter Results from XENON100}

\author{E.~Aprile}\columbia
\author{K.~Arisaka}\losangeles
\author{F.~Arneodo}\assergi
\author{A.~Askin}\zurich
\author{L.~Baudis}\zurich
\author{A.~Behrens}\zurich
\author{K.~Bokeloh}\munster
\author{E.~Brown}\losangeles
\author{T.~Bruch}\zurich
\author{J.~M.~R.~Cardoso}\coimbra
\author{B.~Choi}\columbia
\author{D.~Cline}\losangeles
\author{E.~Duchovni}\weizmann
\author{S.~Fattori}\assergi\mainz
\author{A.~D.~Ferella}\zurich
\author{K.-L.~Giboni}\columbia
\author{E.~Gross}\email{eilam.gross@weizmann.ac.il}\weizmann
\author{A.~Kish}\zurich
\author{C.~W.~Lam}\losangeles
\author{J.~Lamblin}\subatech
\author{R.~F.~Lang}\email{rafael.lang@astro.columbia.edu}\columbia
\author{K.~E.~Lim}\columbia
\author{S.~Lindemann}\heidelberg
\author{M.~Lindner}\heidelberg
\author{J.~A.~M.~Lopes}\coimbra
\author{T.~Marrod\'an~Undagoitia}\zurich
\author{Y.~Mei}\houston
\author{A.~J.~Melgarejo~Fernandez}\columbia
\author{K.~Ni}\shanghai
\author{U.~Oberlack}\mainz\houston
\author{S.~E.~A.~Orrigo}\coimbra
\author{E.~Pantic}\losangeles
\author{G.~Plante}\columbia
\author{A.~C.~C.~Ribeiro}\coimbra
\author{R.~Santorelli}\columbia\zurich
\author{J.~M.~F.~dos Santos}\coimbra
\author{M.~Schumann}\zurich\houston
\author{P.~Shagin}\houston
\author{A.~Teymourian}\losangeles
\author{D.~Thers}\subatech
\author{E.~Tziaferi}\zurich
\author{O.~Vitells}\weizmann
\author{H.~Wang}\losangeles
\author{M.~Weber}\heidelberg
\author{C.~Weinheimer}\munster

\collaboration{The XENON100 Collaboration}\noaffiliation

\begin{abstract}
Many experiments that aim at the direct detection of Dark Matter are able to distinguish a dominant background from the expected feeble signals, based on some measured discrimination parameter. We develop a statistical model for such experiments using the Profile Likelihood ratio as a test statistic in a frequentist approach. We take data from calibrations as control measurements for signal and background, and the method allows the inclusion of data from Monte Carlo simulations. Systematic detector uncertainties, such as uncertainties in the energy scale, as well as astrophysical uncertainties, are included in the model. The statistical model can be used to either set an exclusion limit or to quantify a discovery claim, and the results are derived with a proper treatment of statistical and systematic uncertainties. We apply the model to the first data release of the XENON100 experiment, which allows to extract additional information from the data, and place stronger limits on the spin-independent elastic WIMP-nucleon scattering cross-section. In particular, we derive a single limit, including all relevant systematic uncertainties, with a minimum of $2.4\times10^{-44}\1{cm^2}$ for WIMPs with a mass of $50\1{GeV/c^2}$.
\end{abstract}



\maketitle

%\linenumbers

\section{Introduction}

It is now well established that the dominant mass fraction of our Universe consists of some yet-unknown form of dark matter~\cite{Jarosik:2010iu}. Well-motivated models predict Dark Matter in the form of Weakly Interacting Massive Particles (WIMPs)~\cite{Nakamura:2010zzi} for which searches can be conducted in direct scattering experiments located on Earth. In such experiments, WIMPs are expected to induce nuclear recoils with a roughly exponentially falling spectrum that extends to at most a few tens of keV in energy~\cite{Cerdeno:2010jj}. In contrast, the dominant background is usually composed of electronic recoils from $\beta$- or $\gamma$-radiation. Most direct Dark Matter search experiments are able to discriminate nuclear recoils from electronic recoils based on some discrimination parameter, such as the ratio of ionization to phonon signal~\cite{Ahmed:2009zw,Armengaud:2009hc}, scintillation to phonon signal~\cite{Angloher:2008jj}, ionization to scintillation signal~\cite{Angle:2007uj,Aprile:2010um}, or the pulse shape of the scintillation light~\cite{Lippincott:2008ad}. In a plot of this discrimination parameter as function of energy, the signal and background events thus separate. Typically, a signal acceptance area is defined a priori in this parameter space. Only events that fall within this area are considered as WIMP candidate events for the interpretation of the result in terms of a Dark Matter detection, or for the calculation of limits on the WIMP-nucleon scattering cross-section.

Such a hard cut has some obvious disadvantages. Most importantly, the particular location of the signal candidate events within the signal acceptance area is not taken into account. In general, the expected signal and background will follow a different spectrum, which is utilized by the maximum-gap and optimum-interval methods~\cite{Yellin:2002xd}, a frequentist approach to calculate limits. The generalization to two dimensions, the maximum-patch method~\cite{Henderson:2008bn}, extends this feature to the case where background leaks into the signal acceptance region from high or low discrimination parameters. An intrinsic advantage of these methods is that the resulting limit is robust against small changes of the signal acceptance region, and they are widely used in discriminating direct Dark Matter searches today. However, these methods also have some severe drawbacks: they are designed to always result in an upper limit, and a natural transition to a claim of detection, as for example inherent to the Feldman-Cousins-approach~\cite{Feldman:1997qc}, is not possible. In addition, systematic uncertainties are not taken into account. Thus, the resulting limit can be quoted at a given statistical confidence level~(CL), typically taken to be 90\%, but systematic uncertainties need to be treated separately.

In this paper, we present an approach that overcomes both of these problems of the conventional treatment of data from direct Dark Matter detection experiments. As a practical example, we use data from the first Dark Matter results of the XENON100 experiment~\cite{Aprile:2010um}. This data was previously analyzed in the classical way, by pre-defining a signal acceptance region based on calibration data. In particular, this region was defined using the prompt scintillation signal as a measurement of recoil energy, in the range between 4-20~photoelectrons~(PE). In energy, this corresponds to about $9-33\1{keV_{\n{nr}}}$ (keV nuclear recoil), depending on the energy calibration through the assumed relative scintillation efficiency $\mathcal{L}_{\mathrm{eff}}$~\cite{Hitachi:2005ti,Lindhard:1961zz}. The logarithm of the ratio of proportional scintillation, $S2$, to prompt scintillation, $S1$, was taken as a discrimination parameter~\cite{Aprile:2006kx}, and the signal acceptance region was defined below the median of the nuclear recoil band, as measured with a \ambe~neutron calibration. The electronic recoil discrimination in this signal acceptance region was determined to be larger than 99\% based on calibration data from Compton-scattered $^{60}$Co gammas. Using 11.17~live days of data, no events were observed in the signal acceptance region. Thus, upper limits on the WIMP-nucleon scattering cross-section were derived, based on zero observed events. Two separate limit curves were calculated for two different parameterizations of $\mathcal{L}_{\mathrm{eff}}$ in the low-energy region where measurements are lacking or are uncertain~\cite{Manalaysay:2010mb}.

Here, we re-analyze this data with the Profile Likelihood Ratio statistical approach~\cite{Eadie:1971sm}, that is inherent to the MINOS package commonly used with the MINUIT software~\cite{James:1975dr}. This method relies on the assumption that one can properly model the background via calibration measurements or Monte Carlo simulations, as opposed to the Yellin methods~\cite{Yellin:2002xd}, where no information on the background is assumed. This increases the sensitivity of the method at the price of being less robust against systematic uncertainties on the background. However, by incorporating systematic uncertainties into the model, one can adjust the trade-off between sensitivity and robustness to the level desired by the experimentalist. In any case, understanding of the background is of course essential to any method that is used to quantify the significance of a discovery claim.
%is able to result in a discovery claim.

The particular Profile Likelihood model used here is presented in Section~\ref{sec:pl}. Uncertainties in $\mathcal{L}_{\mathrm{eff}}$, that can serve as an example for systematic uncertainties related to the detector, are treated in Section~\ref{sec:leff}, together with systematic uncertainties from the escape velocity $v_{\n{esc}}$ as an example for astrophysical uncertainties. Calibration data for both the electronic recoil background as well as the nuclear recoil signal are taken as constraining control measurements. These constrain the full likelihood model, which is constructed in Section~\ref{sec:control}. The sensitivity of the experiment is calculated, based solely on this model, in Section~\ref{sec:exclusion}. Taking the actual measurement into account, the resulting exclusion curve is derived also in Section~\ref{sec:exclusion}. Discussion and conclusions follow in Section~\ref{sec:conclusions}.

\section{The Statistical Model}\label{sec:pl}

Events recorded by the XENON100 detector can be characterized by their prompt and proportional scintillation signals, and are therefore treated in this analysis as a set of points in the $(S1,S2)$ plane. The distribution of signal events in $S1$ can be predicted from theoretical models taking into account the detector response, as will be described below. The joint distribution in $S1$ and $S2$, for signal and background events, is estimated from the calibration measurements. The current measurements of $\mathcal{L}_{\mathrm{eff}}$ and $v_{\n{esc}}$ are similarly treated as another set of control measurements. The statistical model describing the outcome of the measurements thus includes a number of unknown quantities, i.e. \textit{nuisance parameters}. Those are the expected number of background events $N_b$ given the exposure and the $(S1,S2)$ parameter range in question, a set of probabilities $\epsilon_s = \{\epsilon_s^j\}$ and $\epsilon_b= \{\epsilon_b^j\}$ describing the distribution of signal and background events in the $(S1,S2)$ plane, the relative scintillation efficiency $\mathcal{L}_{\mathrm{eff}}$, as well as the escape velocity $v_{\n{esc}}$. The full likelihood function $\mathscr{L}$, for a given WIMP mass $m_{\chi}$ and cross-section $\sigma$, is written as a product of five terms:

\begin{eqnarray}\label{eq:likelihood}
\mathscr{L} & = & \mathscr{L}_1(\sigma, N_b, \epsilon_s , \epsilon_b , \mathcal{L}_{\n{eff}} , v_{\n{esc}} ; m_{\chi} )\\
                 &   & \times \mathscr{L}_2(\epsilon_s) \times \mathscr{L}_3(\epsilon_b) \nonumber \\
                 &   & \times \mathscr{L}_4(\mathcal{L}_{\n{eff}}) \times \mathscr{L}_5(v_{\mathrm{esc}}). \nonumber
\end{eqnarray}
The first term $\mathscr{L}_1$ describes the main measurement of the XENON100 detector, while the following terms describe the subsidiary measurements that are used to constrain the nuisance parameters in the main likelihood term $\mathscr{L}_1$. The precise definition of these terms will be given in the following sections. Additional uncertainties can easily be incorporated by adding additional Likelihood terms, but are not relevant for the data set analyzed here.

The signal cross-section $\sigma$ is the one parameter of interest. All other parameters are nuisance parameters which are \textit{profiled out} with a profile likelihood ratio, as explained below (Equation~\ref{eq:pl2}). We follow the procedure of a hypothesis test based on the profile likelihood ratio~\cite{Rolke:2004mj,Cowan:2010js}. This technique can be used both to exclude a WIMP with a specific mass and cross-section, or to establish the significance of a discovery.

\subsection{Exclusion}

A test statistic $q_\sigma$ reduces the observed data to only one value and is constructed in order to test the signal hypothesis $H_\sigma$. It is given by
\begin{eqnarray}\label{eq:pl1}
q_\sigma = \begin{cases} -2\ln\lambda(\sigma) & \;\; \hat \sigma < \sigma \\
                          0                   & \;\; \hat \sigma > \sigma
           \end{cases}
\end{eqnarray}
where $\hat \sigma$ is the maximum likelihood estimator (MLE) of $\sigma$, i.e. the value of $\sigma$ that maximizes the likelihood Equation~\ref{eq:likelihood}. $\lambda(\sigma)$ is the Profile Likelihood ratio and is given by
\begin{eqnarray}\label{eq:pl2}
\lambda(\sigma) &=& \frac{\displaystyle\max_{\sigma ~\mathrm{fixed}}\mathscr{L}\left({ \sigma}; {{\mathcal{L}_{\n{eff}}}}, {v_{\n{esc}}} ,{N}_b,{\epsilon}_s,{\epsilon}_b \right)}{\displaystyle\max \mathscr{L}\left({ \sigma}, {{\mathcal{L}_{\n{eff}}}}, {v_{\n{esc}}} ,{N}_b,{\epsilon}_s,{\epsilon}_b \right)} \nonumber \\
& \equiv & \frac{\mathscr{L} \left( \sigma,\hat{\hat{\mathcal{L}_{\n{eff}}}},\hat{\hat{v_{\n{esc}}}} ,\hat{\hat N}_b,\hat{\hat \epsilon}_s,\hat{\hat\epsilon}_b \right)}
                    {\mathscr{L} \left({\hat \sigma},{\hat{\mathcal{L}_{\n{eff}}}}, \hat{v_{\n{esc}}} ,{\hat N}_b,{\hat \epsilon}_s,{\hat\epsilon}_b \right)}.
\end{eqnarray}
The double-hat parameters in the numerator are the conditional MLEs of the nuisance parameters when the signal cross-section is fixed to a given value $\sigma$. The `single-hat' parameters in the denominator are the MLEs of all parameters allowing also $\sigma$ to vary. By construction, $0 \leq \lambda(\sigma) \leq 1$, hence $q_\sigma \geq 0$. $q_\sigma$ equals zero when the best-fit value of the cross-section ($\hat \sigma$) equals the hypothesized value ($\sigma$), which corresponds to the most signal-like outcome. Larger values of the test statistic $q_\sigma$ indicate that the data are less compatible with the signal hypothesis $H_\sigma$. Since we are concerned with calculating a one-sided upper bound, we only consider outcomes with $\hat \sigma < \sigma$ as an evidence against the signal hypothesis and set $q_\sigma$ to zero otherwise.

Let $f(q_\sigma|H_\sigma)$ be the probability distribution function of the test statistic $q_\sigma$ under the signal hypothesis $H_\sigma$, and let $q^{\n{obs}}_\sigma$ be the value of the test statistic obtained with the observed data. The signal $p$-value $p_s$, is the probability that the outcome of a hypothetical, random XENON100 experiment results in a test statistic larger (less signal-like) than the observed one, when the signal hypothesis $H_\sigma$ is {\it true}. Therefore, $p_s$ given by
\begin{eqnarray}
p_s=\int_{q^{\n{obs}}_\sigma}^\infty f(q_\sigma|H_\sigma) \dd q_\sigma.\label{eq:ps}
\end{eqnarray}
The signal hypothesis $H_\sigma$ is rejected at 90\%~CL if $p_s\leq 10\%$.

Downward fluctuations of the background might lead to exclusions of very small cross-sections to which the experiment is not sensitive. To protect against such an effect, $p_s$ is modified~\cite{Junk:1999kv,Read:2002hq} to
\begin{eqnarray}
p'_s=\frac{p_s}{1-p_b}
\end{eqnarray}
where
\begin{eqnarray}
1-p_b=\int_{q^{\n{obs}}_\sigma}^\infty f(q_\sigma|H_0) \dd q_\sigma
\end{eqnarray}
is the probability of the test statistic $q_\sigma$ to be larger than the observed one under the background-only hypothesis $H_{\sigma=0}\equiv H_0$. Other protection procedures would also be conceivable, but this particular extension has a conservative over-coverage nature. It has been verified by Monte Carlo simulations that the coverage of our claimed 90\%~CL upper bound on the cross-section is in the range (92-95)\% and thus larger than 90\%, in accordance with the conservative nature of the method.

The upper limit $\sigma^{\mathrm{up}}(m_{\chi})$ on the cross-section $\sigma$ for a given WIMP mass
$m_{\chi}$ is found by solving
\begin{eqnarray}
p'_s(\sigma=\sigma^{\mathrm{up}}(m_{\chi}))=10\%.\label{eq:pprime}
\end{eqnarray}

Wilks' theorem~\cite{Wilks:1938ll} states that $q_\sigma$ follows a chi-square distribution in the limit of a large number of observations, which in this context not only includes the observed WIMP candidate events, but also all of the control measurements. It has been verified by Monte Carlo simulations that this asymptotic behavior indeed describes the distribution of the test statistic to a good approximation. We therefore use the chi-square approximation in order to estimate the signal $p$-value $p_s$. $p_b$ is estimated from Monte Carlo simulations of background only events.

\subsection{Discovery}\label{sec:discovery}

%In order to establish a discovery, we can also use the above statistical method in a natural way.
We can also use the above statistical method in a natural way to quantify the significance of a possible additional event population or signal discovery. To this end, we test the background-only hypothesis ($\sigma=0$), and try to reject it. Similar to Equation~\ref{eq:pl1}, the discovery test statistic $q_0$ is defined as
\begin{eqnarray}\label{eq:pl5}
q_0 = \begin{cases} -2\ln\lambda(0) & \;\; \hat \sigma >0 \\
                     0                    & \;\; \hat \sigma <0.
      \end{cases}
\end{eqnarray}
Setting $\sigma=0$ in Equation~\ref{eq:pl2} yields
\begin{eqnarray}\label{eq:pl6}
\lambda(0) = \frac{\mathscr{L} \left(\sigma=0,\hat{\hat{\mathcal{L}_{\n{eff}}}},\hat{\hat{v_{\n{esc}}}},\hat{\hat N}_b,\hat{\hat \epsilon}_s,\hat{\hat\epsilon}_b \right)}
                    {\mathscr{L} \left({\hat \sigma},{\hat{\mathcal{L}_{\n{eff}}}}, \hat{v_{\n{esc}}} ,{\hat N}_b,{\hat \epsilon}_s,{\hat\epsilon}_b \right)}.
\end{eqnarray}
The $p$-value is defined as
\begin{eqnarray}
p_0=\int_{q^{\n{obs}}}^\infty f(q_0|H_0) \dd q_0,
\end{eqnarray}
where $f(q_0|H_0)$ is the probability density function of $q_0$ under the background-only hypothesis $H_0$. A $p$-value of $p_0=0.00135$ corresponds to a 3$\sigma$ signal evidence. Following Wilks' theorem, $q_0$ asymptotically distributes as a chi-square distribution under the background-only hypothesis. Under Wilks' theorem, the discovery significance is given by $Z\sigma$ with $Z=\sqrt{q^{obs}}$. As a general remark, the Profile Likelihood can result in the discovery of a process that looks like the signal hypothesis. However, the interpretation of the physical origin of such a signal is of course left to scientific discourse.

\section{Expected Signal with $\mathcal{L}_{\mathrm{eff}}$ and $v_{\n{esc}}$ as Control Measurements}\label{sec:leff}

\begin{figure}[htbp]
\centering\includegraphics[width=1\columnwidth]{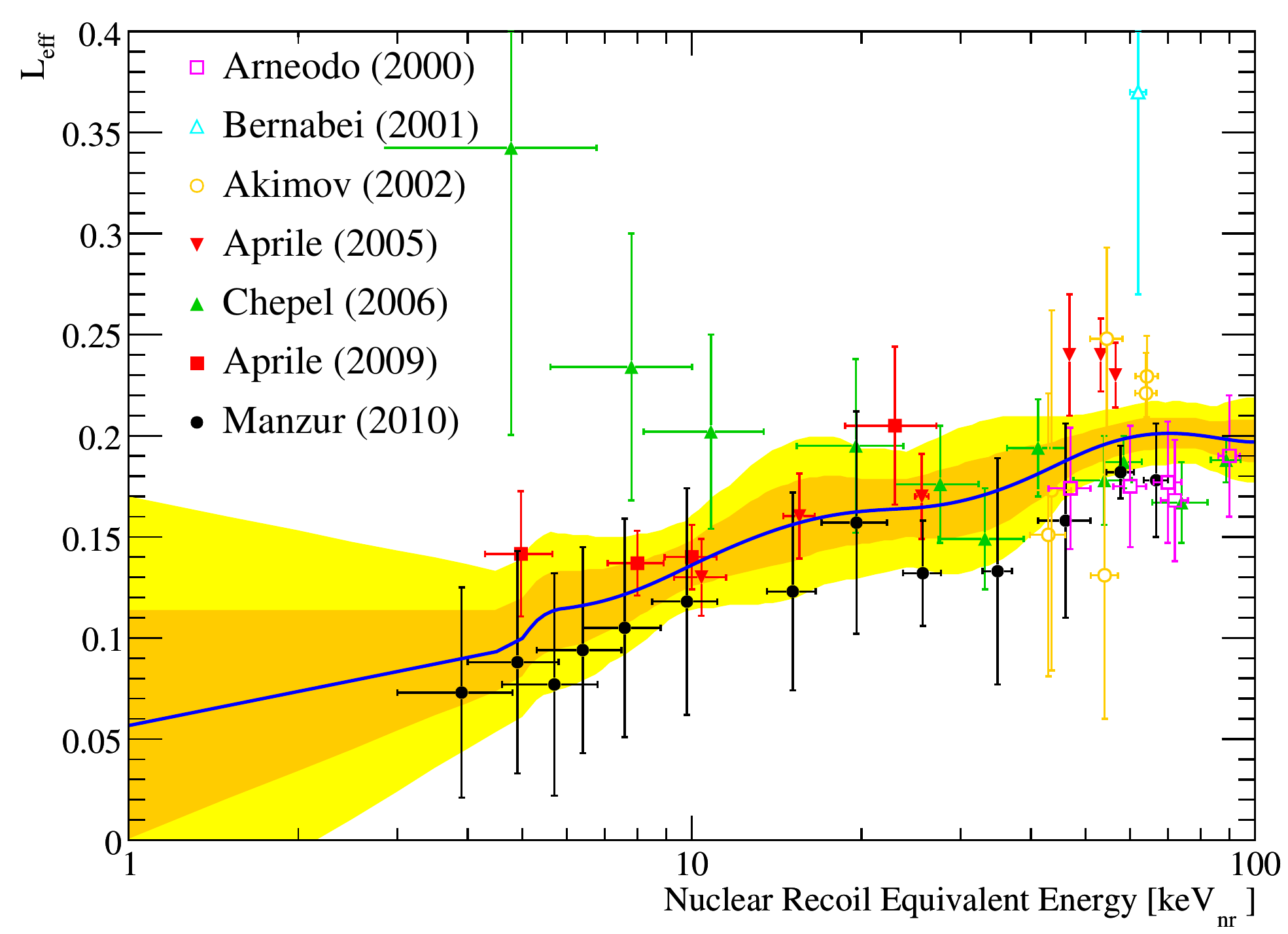}
\caption{All direct measurements of $\mathcal{L}_{\mathrm{eff}}$~\cite{Arneodo:2000vc,Akimov:2001pb,Bernabei:2001pz,Aprile:2005mt,Chepel:2006yv,Aprile:2008rc,Manzur:2009hp} together with a Gaussian fit (solid blue line) and the uncertainty band indicating the $\pm1$ and $\pm2\sigma$ regions (shaded dark and light blue, respectively).}\label{fig:leff}
\end{figure}

The relative scintillation efficiency $\mathcal{L}_{\mathrm{eff}}$ is the largest systematic uncertainty for the data analyzed here. It relates the expected number of $S1$ photoelectrons to the recoil energy $E_{\mathrm{nr}}$. The recoil energy dependence of $\mathcal{L}_{\mathrm{eff}}$ together with its uncertainty are taken from a Gaussian fit of all available direct measurements~\cite{Arneodo:2000vc,Akimov:2001pb,Bernabei:2001pz,Aprile:2005mt,Chepel:2006yv,Aprile:2008rc,Manzur:2009hp}
%indirect: Lebedenko:2008gb, Sorensen:2008ec
as shown in figure~\ref{fig:leff}. To extrapolate below $5\1{keV_{nr}}$, where there are no available measurements of $\mathcal{L}_{\mathrm{eff}}$ yet, the 68\% confidence interval for $\mathcal{L}_{\mathrm{eff}}$ is constructed such that it follows the two different extrapolations used in~\cite{Aprile:2010um} to facilitate the comparison of the resulting limit. We emphasize that this extrapolation follows both the new data published in~\cite{Plante:2011hw}, the theoretical expectations spelled out in~\cite{Bezrukov:2010qa}, as well as the trend expected from simulations~\cite{Szydagis:2011tk}.

We parametrize  the uncertainty of $\mathcal{L}_{\mathrm{eff}}$ with a single nuisance parameter $t$, which is defined to be normally distributed with zero mean and unit variance. Thus, the $\mathcal{L}_{\mathrm{eff}}$-term in Equation~\ref{eq:likelihood} corresponds simply to
\begin{eqnarray}\label{eq:t}
\mathscr{L}_4(\mathcal{L}_{\n{eff}}(t))=\n{exp}(-(t-t_{\n{obs}})^2/2)
\end{eqnarray}
with $t_{\n{obs}}=0$ representing the $\mathcal{L}_{\mathrm{eff}}$-median in Figure~\ref{fig:leff}. $t=-1$ corresponds to the lower boundary of the 68\% confidence region of $\mathcal{L}_{\mathrm{eff}}$ (dark shaded band in Figure~\ref{fig:leff}), etc..

The number $N_s$ of expected signal events is given by the integral over the predicted WIMP energy spectrum $\n{d} R/\n{d}E_{\n{nr}}$~\cite{Lewin:1995rx} as seen in the detector. $\n{d} R/\n{d} E_{\n{nr}}$ is calculated for a given WIMP mass $m_{\chi}$, cross-section $\sigma$, Galactic escape velocity $v_{\n{esc}}$ as well as a given target mass and exposure time.

For the uncertainty on the escape velocity $v_{\n{esc}}$, we use the asymmetric probability distribution $f_v$ as shown in Figure~7 of~\cite{Smith:2006ym}, yielding $498\1{km/s}<v_{\n{esc}}<608\1{km/s}$ at 90\%~CL, with the median being $v_{\n{obs}}=544\1{km/s}$. The likelihood term is defined accordingly to be
\begin{equation}\label{eq:v}
\mathscr{L}_5(v_{esc})= f_v( v_{\n{obs}}|v_{\n{esc}} ).
\end{equation}

Let $\nu(E_{\n{nr}})$ be the expected number of photoelectrons~(PE) for a given recoil energy $E_{\n{nr}}$, given by
\begin{eqnarray}
\nu(E_{\n{nr}}) = E_{\n{nr}} \times \mathcal{L}_{\n{eff}}(E_{\n{nr}}) \times \frac{S_{\n{nr}}}{S_{\n{ee}}} \times L_y.
\end{eqnarray}
$S_{\n{ee}}=0.58$ and $S_{\n{nr}}=0.95$~\cite{Aprile:2006kx} are the scintillation quenching factors due to the electric field for electronic and nuclear recoils, respectively. The normalization light yield $L_y$ for this data sample was measured to be $L_y(122\1{keV_{ee}})=(2.20\pm0.09)\1{PE/keV_{ee}}$~\cite{Aprile:2010um}. Errors on these parameters are small and have been neglected for the current analysis.

The signal rate in number of photoelectrons $n$ is then given by
\begin{eqnarray}
\frac{\n{d} R}{\n{d}n} = \int_0^\infty \n{d}E \; \frac{\n{d} R}{\n{d}E}\left(E;m_{\chi},v_{\mathrm{esc}}\right) \times \n{Poiss}\left( n | \nu(E) \right).
\end{eqnarray}
Taking into account also the finite average single-photoelectron resolution $\sigma_{\n{PMT}}=0.5\1{PE}$ of the XENON100 photomultipliers, the resulting $S1$-spectrum is given by
\begin{eqnarray}\label{eq:rate}
\frac{\n{d} R}{\n{d} S1} & = & \sum_{n=1}^\infty \n{Gauss}( S1 | n ,\sqrt{n}\sigma_{\n{PMT}} ) \nonumber \\
                         &   & \times \frac{\n{d} R}{\n{d}n} \times\zeta_{\n{cuts}}(S1)
\end{eqnarray}
where $\zeta_{\n{cuts}}(S1)$ is the acceptance of the applied cuts~\cite{Aprile:2010um}.

\begin{figure}[!htbp]
\centering\includegraphics[width=1\columnwidth]{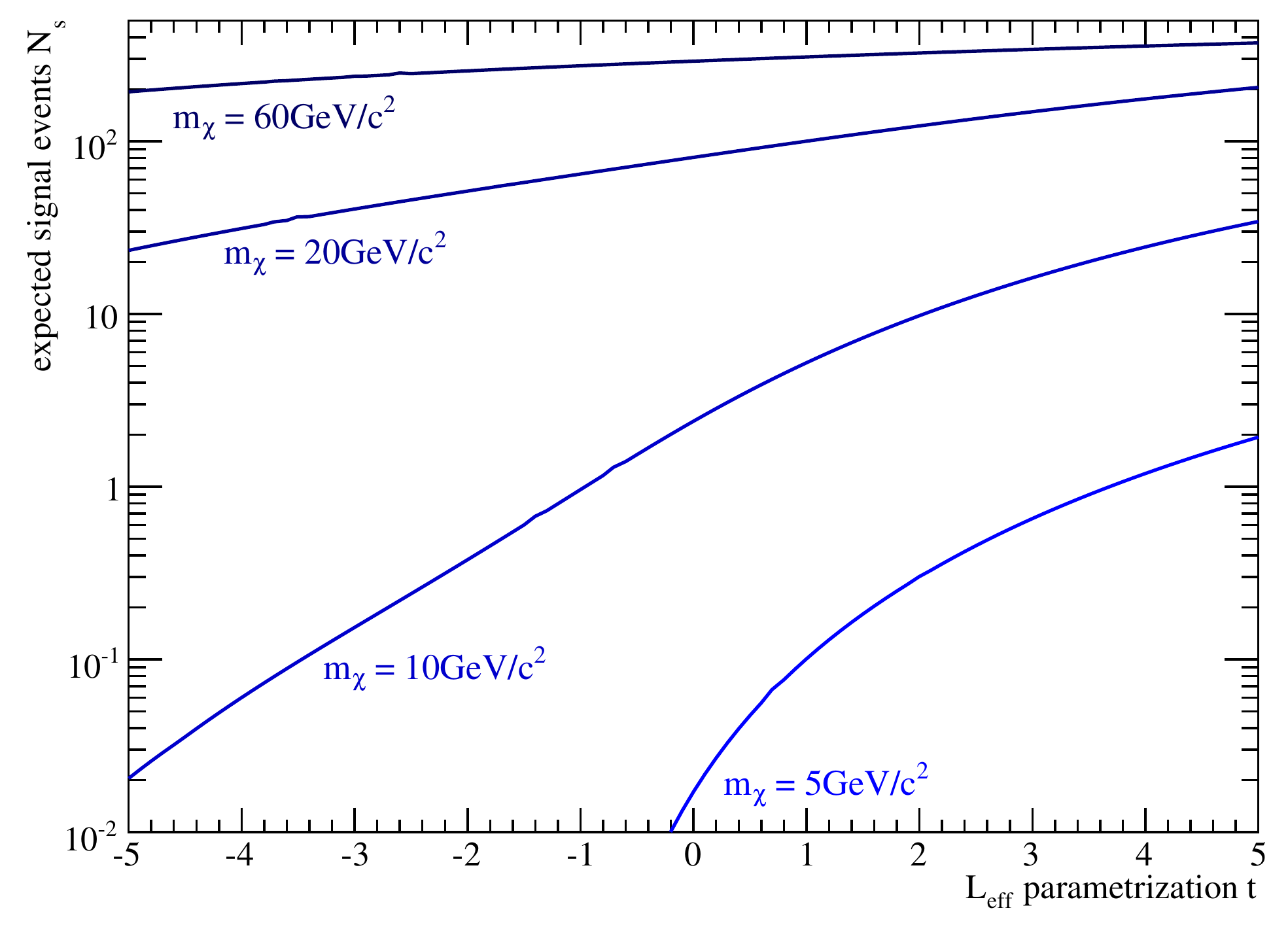}
\caption{Expected number of signal events $N_s$ as a function of the $\mathcal{L}_{\mathrm{eff}}$ parametrization $t$, for various WIMP masses $m_{\chi}$ as indicated in the figure, and for a cross-section $\sigma=10^{-39}\1{cm^2}$. The lower the WIMP mass $m_{\chi}$, the stronger the systematic uncertainty from $\mathcal{L}_{\mathrm{eff}}$.}\label{fig:ns}
\end{figure}

The total number of expected signal events $N_s(t,v_{\n{esc}},\sigma)$ is calculated from the integral
\begin{eqnarray}
N_s(t,v_{\n{esc}},\sigma) = \int_{S1_{\n{lower}}}^{S1_{\n{upper}}} \frac{\n{d}R}{\n{d}S1}(m_{\chi})\dd S1
\end{eqnarray}
where $S1_{\n{lower}}=4\1{PE}$ and $S1_{\n{upper}}=20\1{PE}$ is the energy interval considered in the analysis. The effect of varying $t$~(Equation~\ref{eq:t}) for different WIMP masses is shown in Figure~\ref{fig:ns}. It is calculated for an exposure of 11.17~days with 40~kg of xenon and a cross-section of $\sigma=10^{-39}\1{cm^2}$.

\begin{figure}[htbp]
\centering\includegraphics[width=1\columnwidth]{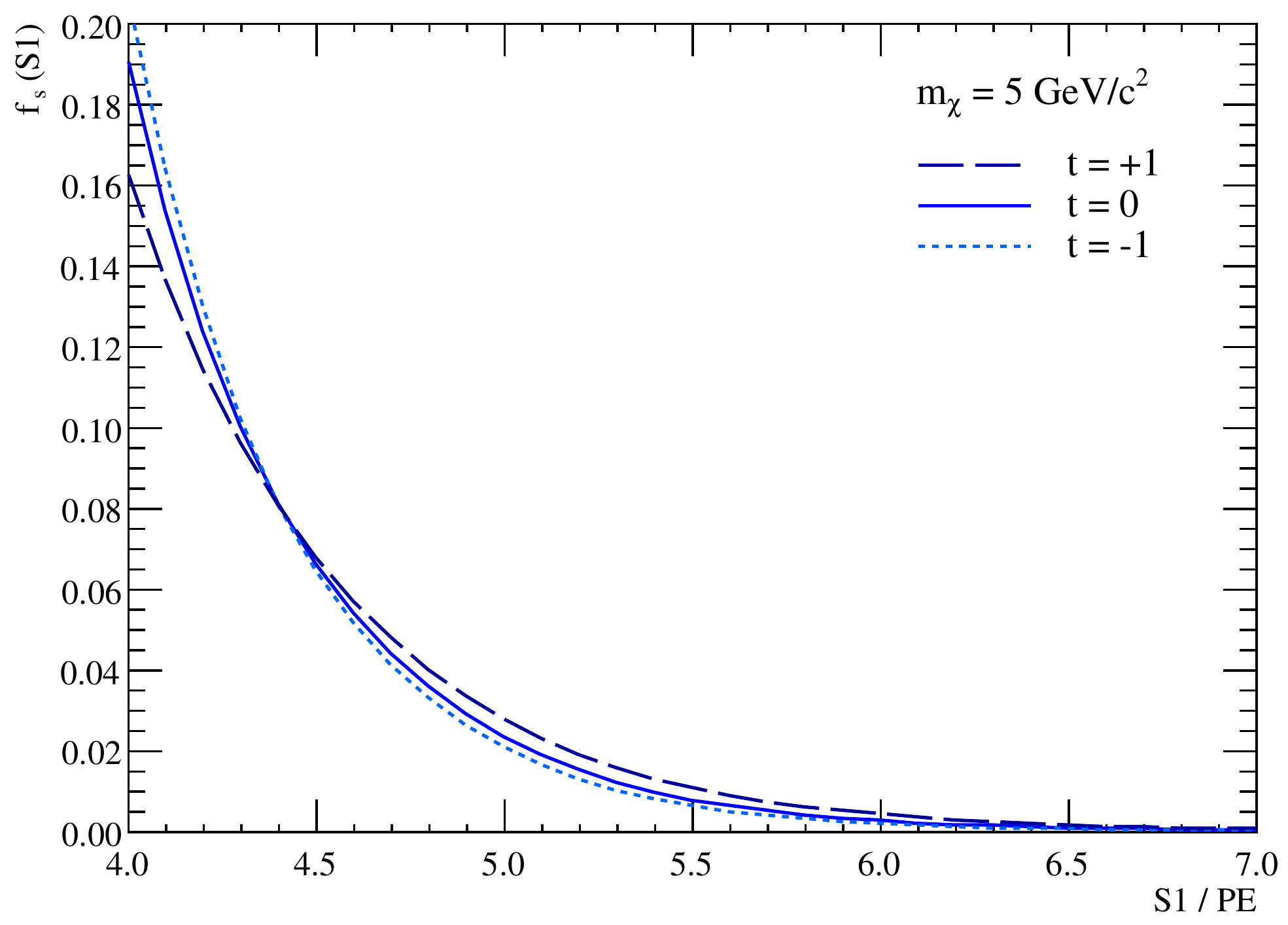}
\caption{The normalized WIMP recoil spectrum $f_s(S1)$ for $m_{\chi}=5\1{GeV/c^2}$ and three possible parameterizations of $\mathcal{L}_{\mathrm{eff}}$ through the parameter $t$~(Equation~\ref{eq:t}). Even for low WIMP masses, the spectral shape hardly changes.}\label{fig:fsofs1}
\end{figure}

Finally, the predicted normalized WIMP spectrum $f_s(S1)$ is given by
\begin{eqnarray}\label{eq:fs}
f_s(S1 ; m_{\chi}) = \frac{ \displaystyle\frac{\n{d}R}{\n{d}S1}(m_{\chi}) }{ \displaystyle \int_{S1_{\n{lower}}}^{S1_{\n{upper}}} \frac{\n{d}R}{\n{d}S1}(m_{\chi}) \dd S1 }.
\end{eqnarray}
Variation of $t$ modifies the expected number of total signal events at a given cross-section, but the shape of the $S1$ distribution is found to be hardly affected, see Figure~\ref{fig:fsofs1}. We therefore assume that $f_s$ is independent of $t$, which greatly simplifies the analysis.

\section{Likelihood Construction}\label{sec:control}

\begin{figure}[htbp]
\centering\includegraphics[width=1\columnwidth]{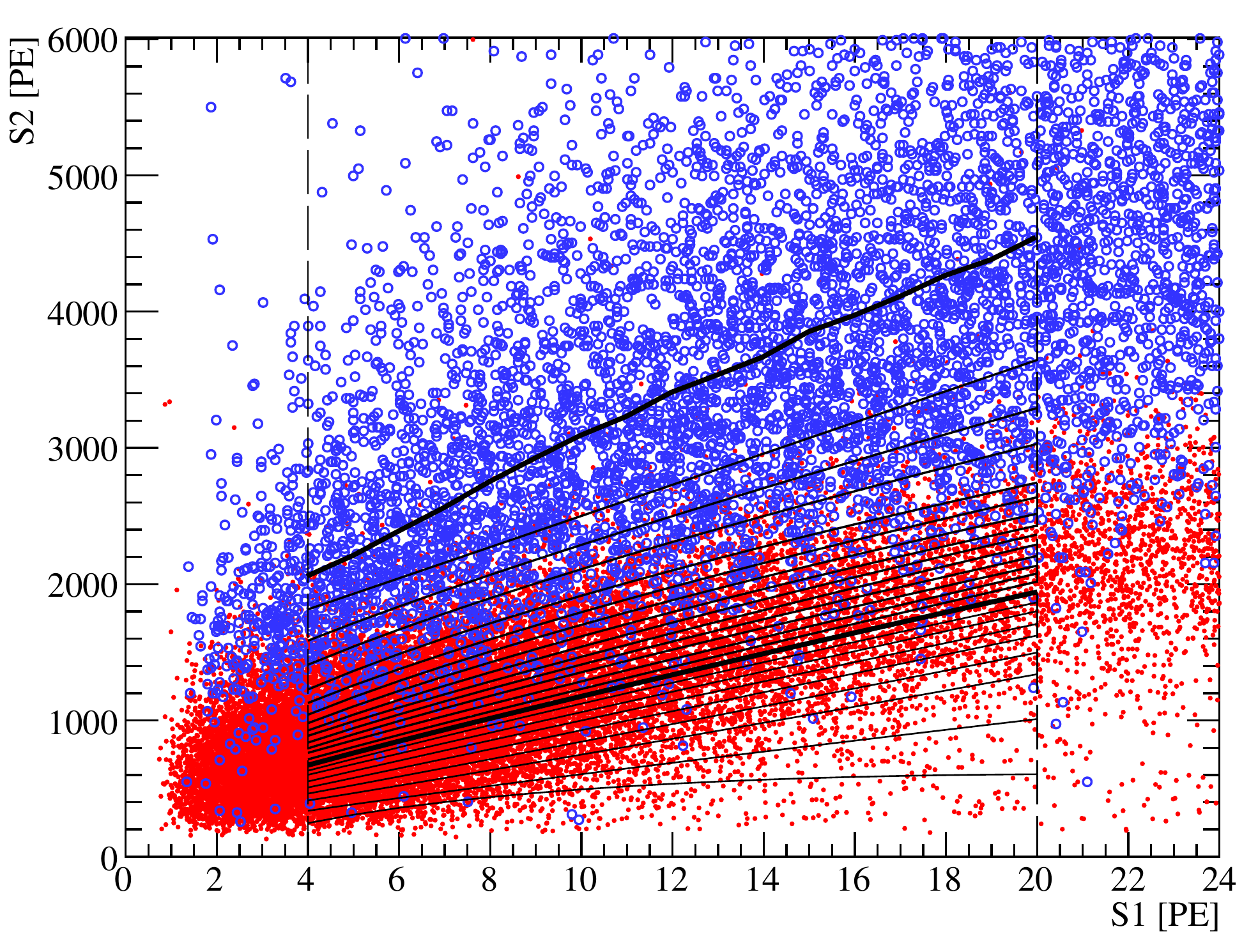}
\caption{Scatter plot of electronic recoils from a $^{60}$Co source (blue circles) and nuclear recoils from an \ambe~source (red dots). Superimposed are the border lines of the bands $j$ (thin lines) as well as the median of the electronic (upper thick line) and nuclear (lower thick line) recoil bands in the relevant $S1$ parameter range (dashed vertical lines).}\label{fig:bands}
\end{figure}

The total number of calibration events used for this analysis in the 4~PE-20~PE $S1$-region was $M_s=18907$ nuclear recoils from \ambe~and $M_b=5053$ electronic recoils from Compton-scattered $^{60}$Co gammas. Those were divided into $j=1,\ldots,(K=23)$ bands in the $(S1,S2)$ plane, containing approximately equal fractions (about 4\%) of nuclear recoils in each band, see Figure~\ref{fig:bands}. $\epsilon_s^j$ and $\epsilon_b^j$ are the corresponding probabilities for a signal or background event to fall in a given band $j$. These bands were constructed such that the fraction of nuclear recoil events in each band is independent of $S1$, so that the signal $S1$ spectrum $f_s$ is the same in all bands. The binning resolution of the $(S1,S2)$-plane is limited by the amount of available neutron calibration data. It was verified that the resulting limit is not sensitive to the number of bands.

We take the calibration measurements as control measurements of $m_s^j$ and $m_b^j$ events in band $j$, in order to constrain $\epsilon_s^j$ and $\epsilon_b^j$. $m_s^j$ and $m_b^j$ are assumed to be Poisson distributed with expectations $\epsilon_s^j M_s$ and $\epsilon_b^j M_b$. The corresponding likelihood terms in Equation~\ref{eq:likelihood} are thus
\begin{eqnarray}
\mathscr{L}_2(\epsilon_s^j) & = & \prod_j^K \n{Poiss}(m_s^j |\epsilon^j_s M_s), \\
\mathscr{L}_3(\epsilon_b^j) & = & \prod_j^K \n{Poiss}(m_b^j |\epsilon^j_b M_b).
\end{eqnarray}

Finally, given a set of $n^j$ data points $(S1_i,S2_i)$ in a band $j$, the likelihood that they emerge from a given WIMP spectrum is given by
\begin{eqnarray}
\mathscr{L}_1 & = & \prod_j^K \n{Poiss}(n^j | \epsilon^j_s N_s+\epsilon^j_b N_b) \nonumber \\
              &   & \times \prod_{i=1}^{n^j} \frac{\epsilon^j_s N_s f_s(S1_i)+\epsilon^j_b N_b f_b(S1_i)}{ \epsilon^j_s N_s + \epsilon^j_b N_b}
\end{eqnarray}
where $f_s$ is given by Equation~\ref{eq:fs}. In accordance with both observation and Monte Carlo simulations, the electronic recoil background spectrum $f_b$ is assumed to be flat in energy~\cite{Aprile:2011vb}.

Collecting all the likelihood terms together, the full likelihood function reads
\begin{eqnarray}\label{eq:lhd}
\mathscr{L} & = & \prod_{j=1}^{K} \n{Poiss}(n^j | \epsilon^j_s N_s +\epsilon^j_b N_b) \nonumber  \\
            &   & \times \prod_{i=1}^{n^j} \frac{\epsilon^j_s N_s f_s(S1)+\epsilon^j_b N_b f_b(S1)}{ \epsilon^j_s N_s+ \epsilon^j_b N_b} \nonumber \\
            &   & \times \n{Poiss}(m_b^j | \epsilon^j_b M_b) \times \n{Poiss}(m_s^j |\epsilon^j_s M_s) \nonumber \\
            &   & \times e^{-(t-t^{\n{obs}})^2/2} \times f_v(v_{\n{obs}}|v_{\n{esc}}).
\end{eqnarray}
Differentiating $ \log\mathscr{L}$ with respect to the expected number of signal events $N_s$ and with respect to $\epsilon_s^j,\epsilon_b^j$, we find the following relations between the MLEs:
\begin{eqnarray}
N \equiv \hat N_s + \hat N_b & = & \sum_{j=1}^{K} n^j \\
(N_s + M_s) \hat \epsilon^j_s + (N_b + M_b)\hat \epsilon^j_b & = & n^j +m_s^j + m_b^j.
\end{eqnarray}
These relations prove useful in the minimization of the likelihood.

\section{WIMP Exclusion}\label{sec:exclusion}

\begin{figure}[htbp]
\centering\includegraphics[width=1\columnwidth]{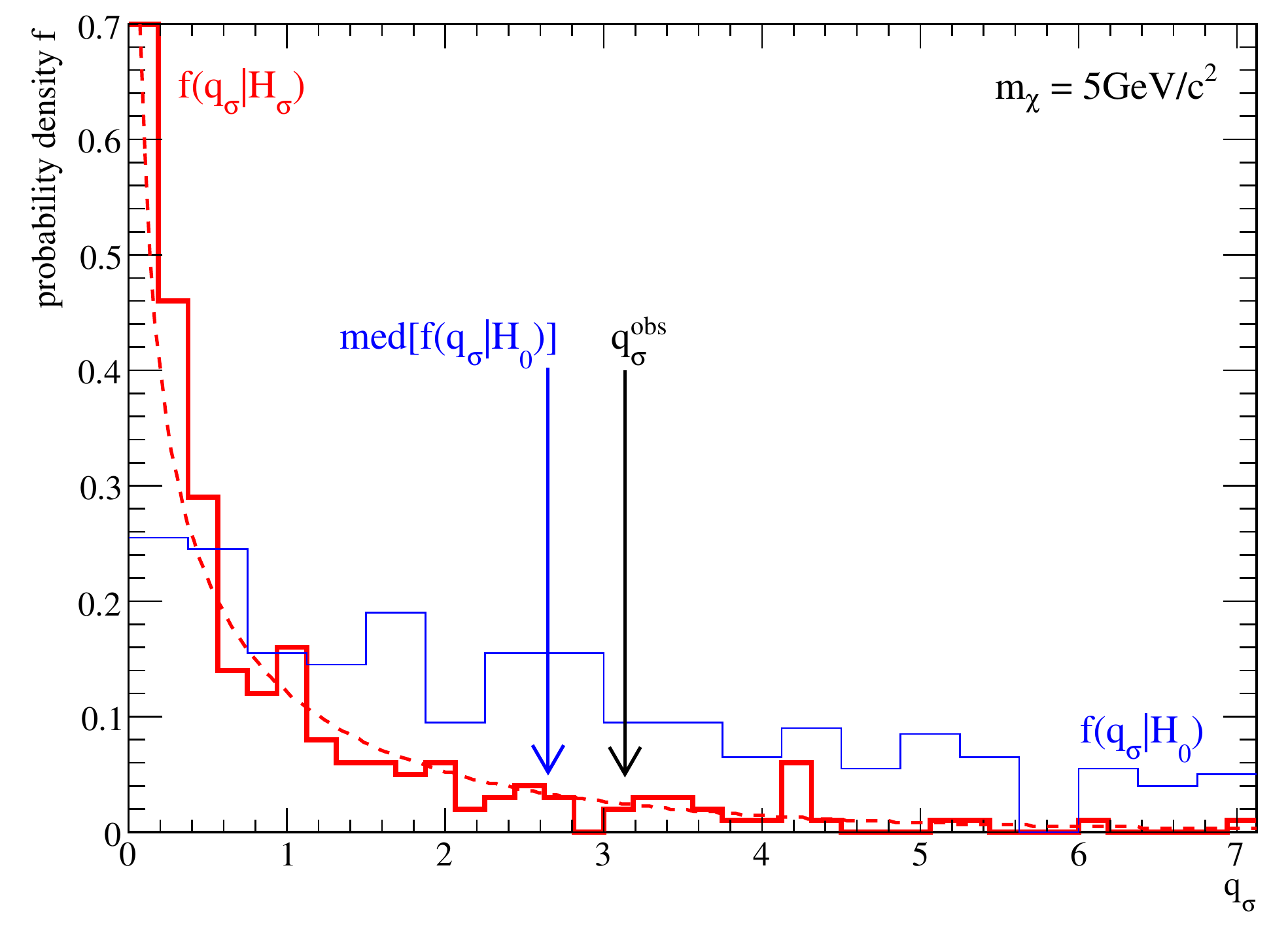}
\caption{The probability density functions $f(q_\sigma|H_\sigma)$ (thick red histogram) and $f(q_\sigma|H_0)$ (thin blue histogram) for a WIMP with $m_{\chi}=5\1{GeV/c^2}$ and $\sigma = \sigma^{up}$. One can clearly see the distribution is well approximated by a chi-square distribution (dashed red line). The median value of $f(q_\sigma|H_0)$ is indicated together with the test statistic $q^{\n{obs}}_\sigma$ observed in data.}\label{fig:fofq}
\end{figure}

To assess the statistical power of the experiment, the exclusion sensitivity can be calculated even before analyzing the actual data. The sensitivity of the experiment is the limit that can be expected for given exposure and experimental conditions. The test statistic $q_\sigma$ is obtained by plugging the likelihood (Equation~\ref{eq:lhd}) into equations~\ref{eq:pl1} and~\ref{eq:pl2}. To verify the validity of Wilks' theorem, we generate a signal for a given WIMP mass and add it to a background simulation. The resulting test statistic distribution $f(q_\sigma|H_\sigma)$ for a $5\1{GeV/c^2}$ WIMP is shown in Figure~\ref{fig:fofq} (thick red histogram). One can clearly see that the distribution is well approximated by a chi-square distribution. This allows to estimate $p_s$ analytically.

\begin{figure}[htbp]
\centering\includegraphics[width=1\columnwidth]{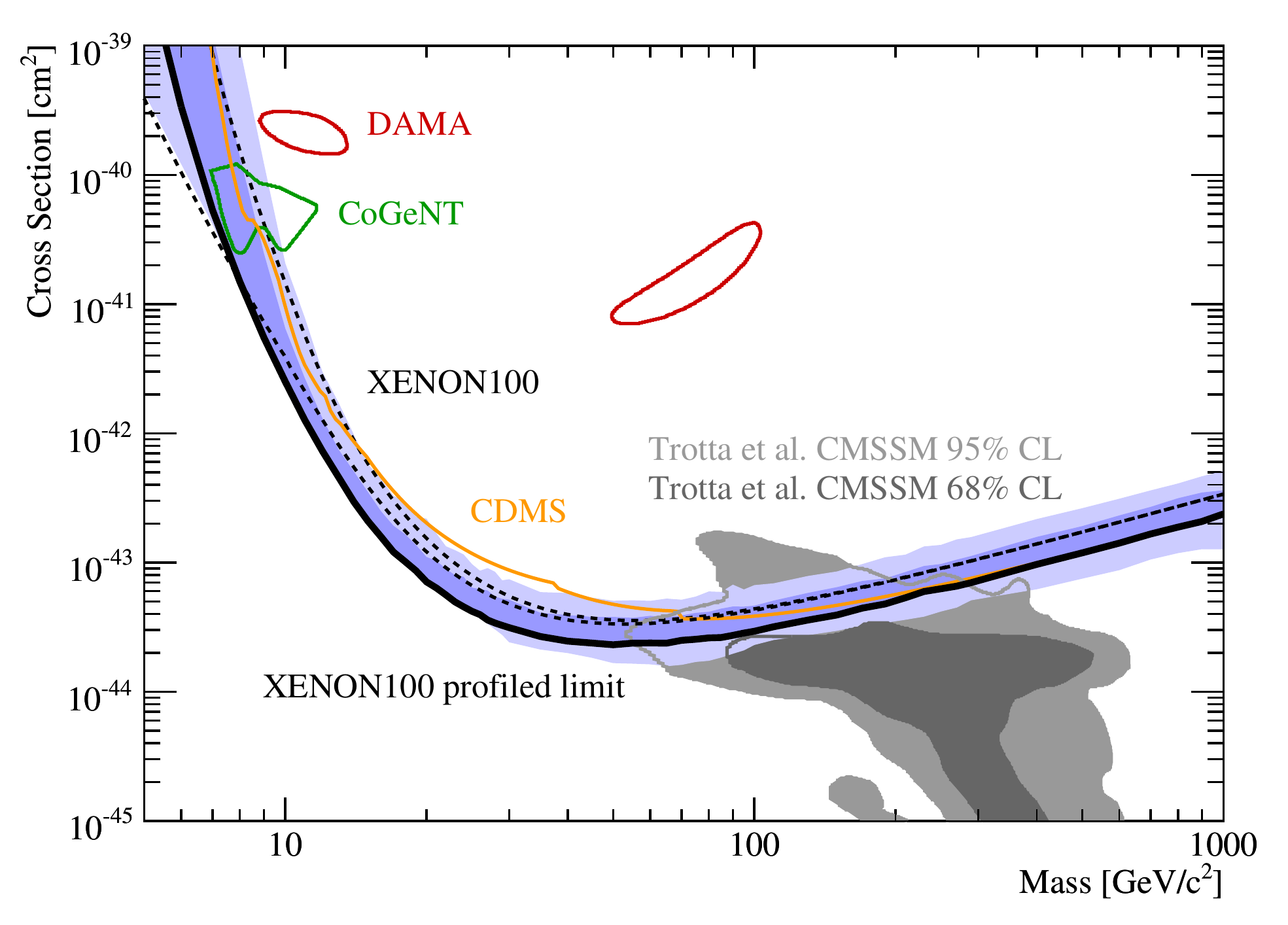}
\caption{Parameter space of spin-independent elastic WIMP-nucleon cross-section $\sigma$ as function of WIMP mass $m_{\chi}$. The sensitivity for the data set analyzed here is shown as light and dark (blue) shaded areas at $1\sigma$ and $2\sigma$~CL, respectively. The actual limit at 90\%~CL, taking into account all relevant systematic uncertainties as derived with the Profile Likelihood method, is shown as the thick (black) line. Two limits from the same data set, derived for two assumptions of the behavior of $\mathcal{L}_{\mathrm{eff}}$, are shown as dotted lines~\cite{Aprile:2010um}. A limit from CDMS~\cite{Ahmed:2009zw} is shown as thin (orange) line, re-calculated assuming an escape velocity of $544\1{km/s}$ and $v_0 = 220\1{km/s}$. Expectations from a theoretical model~\cite{Trotta:2008bp} are shown, as well as the areas (at $90\percent$ CL) favored by CoGeNT (green)~\cite{Aalseth:2010vx} and DAMA (red, without channeling)~\cite{Savage:2008er}.}\label{fig:limit}
\end{figure}

To estimate the exclusion sensitivity, background-only experiments are simulated, based on a Poisson distribution with the expectation of 22~events as observed in~\cite{Aprile:2010um}. The test statistic distribution $f(q_\sigma|H_0)$ is shown as the thin (blue) curve in Figure~\ref{fig:fofq}. The sensitivity is defined as the median $\n{med}[q_\sigma]$ of $f(q_\sigma|H_0)$, also indicated in the figure. One can see that a large fraction of experiments under the background-only hypothesis (thin blue histogram) result in a test statistic $q_\sigma$ that is similar to or even more signal-like (the area to the left of $q_\sigma^{obs}$) than that of experiments with a signal (thick red histogram). Figure~\ref{fig:limit} shows the sensitivity as its $1\sigma$ (dark shaded) and $2\sigma$ (light shaded) bands. For completeness of the statistical analysis we note that the $p$-value of the background-only hypothesis is 50\% as there are less events than expected, $\hat \sigma<0$ and $q_0=0$.

To derive the upper bound on the WIMP cross-section as function of WIMP mass, we solve Equation~\ref{eq:pprime} for the cross-section $\sigma=\sigma^{\mathrm{up}}(m_{\chi})$ that satisfies $p'_s=10\%$ (see Section~\ref{sec:pl}). The test statistic $q_\sigma^{\n{obs}}$ actually observed in data is shown in Figure~\ref{fig:fofq}, and the actual limit is shown in Figure~\ref{fig:limit}. This limit has a minimum of $\sigma^{\n{up}} = 2.4\times10^{-44}\1{cm^2}$ at $m_{\chi}\sim50\1{GeV/c^2}$.

Two major advantages of the method presented here manifest themselves when comparing this new limit to the conventionally calculated one previously published in~\cite{Aprile:2010um}. In the high WIMP mass region, the limit obtained here is stronger by about 20\%-30\%. This simply reflects the fact that in~\cite{Aprile:2010um}, only half of the available parameter space (50\% nuclear recoil acceptance) was defined a priori as the signal acceptance region. In contrast, the current analysis considers the full available discrimination parameter space, also taking the background in each band $j$ into account. Since there are no events observed in the data even in the $1\sigma$ region above the nuclear recoil median (see Figure~3 in~\cite{Aprile:2010um}), the limit improves accordingly.

The second major difference between these two analyses appears in the low WIMP mass region, which is most sensitive to uncertainties in $\mathcal{L}_{\mathrm{eff}}$. The limit obtained here is more stringent than the limit shown in~\cite{Aprile:2010um}, constraining the areas favored by CoGeNT~\cite{Aalseth:2010vx} and DAMA~\cite{Bernabei:2010mq,Savage:2008er}. This comes from our definition of the uncertainty bands of $\mathcal{L}_{\mathrm{eff}}$ as shown in Figure~\ref{fig:leff}, where we set the $1\sigma$ interval to equal the two extrapolations used in the previous analysis. We emphasize that here we do not assume any specific behavior of $\mathcal{L}_{\mathrm{eff}}$, but rather consider it as an unknown quantity with the appropriate measurement uncertainty, as described in Section~\ref{sec:leff}. Such systematic uncertainty is incorporated in a natural way in the calculation of the limit and results in a single limit at 90\%~CL, including all considered uncertainties.

\section{Conclusions}\label{sec:conclusions}

We have introduced and applied the frequentist statistical method based on the profile likelihood test statistic to re-analyze the first data release from the XENON100 direct Dark Matter search experiment. This method avoids the need to a priori define a signal acceptance region, but instead takes all measured data into account. The background was estimated using calibration data as control measurements. Uncertainties in the relative scintillation efficiency $\mathcal{L}_{\mathrm{eff}}$ and the Galactic escape velocity $v_{\mathrm{esc}}$ were taken into account in the construction of the likelihood model. Using the profile likelihood test statistic allows to calculate the sensitivity of the experiment with its uncertainty bands, and to set a well-defined single limit, taking all systematic uncertainties into account. Applying the method to the previously published XENON100 data results in an improvement of the limit over a wide WIMP mass range, with a minimum $\sigma^{\n{up}}<2.4\times10^{-44}\1{cm^2}$ for WIMPs with mass $m_{\chi}=50\1{GeV/c^2}$. In addition, this method can easily be applied for the discovery of a WIMP signal.

\vspace{1cm}

\section{Acknowledgements}

We gratefully acknowledge support from NSF Grants No.~PHY07-05326, PHY07-05337, PHY09-04220, PHY09-04212, and PHY09-04224, DOE Grant No.~DE-FG-03-91ER40662, SNF Grants No.~20-118119 and 20-126993, the Volkswagen Foundation, FCT Grant No.~PTDC/FIS/100474/2008, STCSM Grant No.~10ZR1415000, the Minerva Gesellschaft and GIF. We are grateful to the LNGS staff for their continued support.

%\bibliographystyle{apsrev}
%\bibliography{xenonprofile}

\begin{thebibliography}{37}
\expandafter\ifx\csname natexlab\endcsname\relax\def\natexlab#1{#1}\fi
\expandafter\ifx\csname bibnamefont\endcsname\relax
  \def\bibnamefont#1{#1}\fi
\expandafter\ifx\csname bibfnamefont\endcsname\relax
  \def\bibfnamefont#1{#1}\fi
\expandafter\ifx\csname citenamefont\endcsname\relax
  \def\citenamefont#1{#1}\fi
\expandafter\ifx\csname url\endcsname\relax
  \def\url#1{\texttt{#1}}\fi
\expandafter\ifx\csname urlprefix\endcsname\relax\def\urlprefix{URL }\fi
\providecommand{\bibinfo}[2]{#2}
\providecommand{\arxiv}[2][]{\url{#2}}

\bibitem[{\citenamefont{Jarosik et~al.}(2011)}]{Jarosik:2010iu}
\bibinfo{author}{\bibfnamefont{N.}~\bibnamefont{Jarosik}} \bibnamefont{et~al.},
  \bibinfo{journal}{Astrophys. J. Suppl.} \textbf{\bibinfo{volume}{192}},
  \bibinfo{pages}{14} (\bibinfo{year}{2011}), \arxiv{1001.4744}.

\bibitem[{\citenamefont{Nakamura et~al.}(2010)}]{Nakamura:2010zzi}
\bibinfo{author}{\bibfnamefont{K.}~\bibnamefont{Nakamura}} \bibnamefont{et~al.}
  (\bibinfo{collaboration}{Particle Data Group}), \bibinfo{journal}{J. Phys.}
  \textbf{\bibinfo{volume}{G37}}, \bibinfo{pages}{075021}
  (\bibinfo{year}{2010}).

\bibitem[{\citenamefont{Cerdeno and Green}(2010)}]{Cerdeno:2010jj}
\bibinfo{author}{\bibfnamefont{D.~G.} \bibnamefont{Cerdeno}} \bibnamefont{and}
  \bibinfo{author}{\bibfnamefont{A.~M.} \bibnamefont{Green}},
  \textit{\bibinfo{title}{Direct detection of WIMPs}}
  (\bibinfo{publisher}{Cambridge University Press}, \bibinfo{year}{2010}),
  \bibinfo{note}{{Chapter 17 in: \textit{Particle Dark Matter: Observations,
  Models and Searches}, ed. G. Bertone}}, \arxiv{1002.1912}.

\bibitem[{\citenamefont{Ahmed et~al.}(2010)}]{Ahmed:2009zw}
\bibinfo{author}{\bibfnamefont{Z.}~\bibnamefont{Ahmed}} \bibnamefont{et~al.},
  \bibinfo{journal}{Science} \textbf{\bibinfo{volume}{327}},
  \bibinfo{pages}{1619} (\bibinfo{year}{2010}), \arxiv{0912.3592}.

\bibitem[{\citenamefont{Armengaud et~al.}(2010)}]{Armengaud:2009hc}
\bibinfo{author}{\bibfnamefont{E.}~\bibnamefont{Armengaud}}
  \bibnamefont{et~al.}, \bibinfo{journal}{Phys. Lett.}
  \textbf{\bibinfo{volume}{B687}}, \bibinfo{pages}{294} (\bibinfo{year}{2010}),
  \arxiv{0912.0805}.

\bibitem[{\citenamefont{Angloher et~al.}(2009)}]{Angloher:2008jj}
\bibinfo{author}{\bibfnamefont{G.}~\bibnamefont{Angloher}}
  \bibnamefont{et~al.}, \bibinfo{journal}{Astropart. Phys.}
  \textbf{\bibinfo{volume}{31}}, \bibinfo{pages}{270} (\bibinfo{year}{2009}),
  \arxiv{0809.1829}.

\bibitem[{\citenamefont{Angle et~al.}(2008)}]{Angle:2007uj}
\bibinfo{author}{\bibfnamefont{J.}~\bibnamefont{Angle}} \bibnamefont{et~al.},
  \bibinfo{journal}{Phys. Rev. Lett.} \textbf{\bibinfo{volume}{100}},
  \bibinfo{pages}{021303} (\bibinfo{year}{2008}), \arxiv{0706.0039}.

\bibitem[{\citenamefont{Aprile et~al.}(2010)}]{Aprile:2010um}
\bibinfo{author}{\bibfnamefont{E.}~\bibnamefont{Aprile}} \bibnamefont{et~al.},
  \bibinfo{journal}{Phys. Rev. Lett.} \textbf{\bibinfo{volume}{105}},
  \bibinfo{pages}{131302} (\bibinfo{year}{2010}), \arxiv{1005.0380}.

\bibitem[{\citenamefont{Lippincott et~al.}(2008)}]{Lippincott:2008ad}
\bibinfo{author}{\bibfnamefont{W.~H.} \bibnamefont{Lippincott}}
  \bibnamefont{et~al.}, \bibinfo{journal}{Phys. Rev.}
  \textbf{\bibinfo{volume}{C78}}, \bibinfo{pages}{035801}
  (\bibinfo{year}{2008}), \arxiv{0801.1531}.

\bibitem[{\citenamefont{Yellin}(2002)}]{Yellin:2002xd}
\bibinfo{author}{\bibfnamefont{S.}~\bibnamefont{Yellin}},
  \bibinfo{journal}{Phys. Rev.} \textbf{\bibinfo{volume}{D66}},
  \bibinfo{pages}{032005} (\bibinfo{year}{2002}), \arxiv{physics/0203002},
  \bibinfo{note}{{See also
  S.~Yellin (2007), \textit{Extending the optimum interval method},
  \arxiv{0709.2701}}}.

\bibitem[{\citenamefont{Henderson et~al.}(2008)\citenamefont{Henderson, Monroe,
  and Fisher}}]{Henderson:2008bn}
\bibinfo{author}{\bibfnamefont{S.}~\bibnamefont{Henderson}},
  \bibinfo{author}{\bibfnamefont{J.}~\bibnamefont{Monroe}}, \bibnamefont{and}
  \bibinfo{author}{\bibfnamefont{P.}~\bibnamefont{Fisher}},
  \bibinfo{journal}{Phys. Rev.} \textbf{\bibinfo{volume}{D78}},
  \bibinfo{pages}{015020} (\bibinfo{year}{2008}), \arxiv{0801.1624}.

\bibitem[{\citenamefont{Feldman and Cousins}(1998)}]{Feldman:1997qc}
\bibinfo{author}{\bibfnamefont{G.~J.} \bibnamefont{Feldman}} \bibnamefont{and}
  \bibinfo{author}{\bibfnamefont{R.~D.} \bibnamefont{Cousins}},
  \bibinfo{journal}{Phys. Rev.} \textbf{\bibinfo{volume}{D57}},
  \bibinfo{pages}{3873} (\bibinfo{year}{1998}), \arxiv{physics/9711021}.

\bibitem[{\citenamefont{Hitachi}(2005)}]{Hitachi:2005ti}
\bibinfo{author}{\bibfnamefont{A.}~\bibnamefont{Hitachi}},
  \bibinfo{journal}{Astropart. Phys.} \textbf{\bibinfo{volume}{24}},
  \bibinfo{pages}{247} (\bibinfo{year}{2005}).

\bibitem[{\citenamefont{Lindhard and Scharff}(1961)}]{Lindhard:1961zz}
\bibinfo{author}{\bibfnamefont{J.}~\bibnamefont{Lindhard}} \bibnamefont{and}
  \bibinfo{author}{\bibfnamefont{M.}~\bibnamefont{Scharff}},
  \bibinfo{journal}{Phys. Rev.} \textbf{\bibinfo{volume}{124}},
  \bibinfo{pages}{128} (\bibinfo{year}{1961}).

\bibitem[{\citenamefont{Aprile et~al.}(2006)}]{Aprile:2006kx}
\bibinfo{author}{\bibfnamefont{E.}~\bibnamefont{Aprile}} \bibnamefont{et~al.},
  \bibinfo{journal}{Phys. Rev. Lett.} \textbf{\bibinfo{volume}{97}},
  \bibinfo{pages}{081302} (\bibinfo{year}{2006}), \arxiv{astro-ph/0601552}.

\bibitem[{\citenamefont{Manalaysay}(2010)}]{Manalaysay:2010mb}
\bibinfo{author}{\bibfnamefont{A.}~\bibnamefont{Manalaysay}}
  (\bibinfo{year}{2010}), \arxiv{1007.3746}.

\bibitem[{\citenamefont{Eadie et~al.}(1971)\citenamefont{Eadie, Drijard, James,
  Roos, and Sadoulet}}]{Eadie:1971sm}
\bibinfo{author}{\bibfnamefont{W.}~\bibnamefont{Eadie}},
  \bibinfo{author}{\bibfnamefont{D.}~\bibnamefont{Drijard}},
  \bibinfo{author}{\bibfnamefont{F.}~\bibnamefont{James}},
  \bibinfo{author}{\bibfnamefont{M.}~\bibnamefont{Roos}}, \bibnamefont{and}
  \bibinfo{author}{\bibfnamefont{B.}~\bibnamefont{Sadoulet}},
  \textit{\bibinfo{title}{Statistical Methods in Experimental Physics}}
  (\bibinfo{publisher}{North-Holland Publ. Co., Amsterdam},
  \bibinfo{year}{1971}), \bibinfo{note}{{see in particular pp. 203--205}}.

\bibitem[{\citenamefont{James and Roos}(1975)}]{James:1975dr}
\bibinfo{author}{\bibfnamefont{F.}~\bibnamefont{James}} \bibnamefont{and}
  \bibinfo{author}{\bibfnamefont{M.}~\bibnamefont{Roos}},
  \bibinfo{journal}{Comput. Phys. Commun.} \textbf{\bibinfo{volume}{10}},
  \bibinfo{pages}{343} (\bibinfo{year}{1975}).

\bibitem[{\citenamefont{Rolke et~al.}(2005)\citenamefont{Rolke, Lopez, and
  Conrad}}]{Rolke:2004mj}
\bibinfo{author}{\bibfnamefont{W.~A.} \bibnamefont{Rolke}},
  \bibinfo{author}{\bibfnamefont{A.~M.} \bibnamefont{Lopez}}, \bibnamefont{and}
  \bibinfo{author}{\bibfnamefont{J.}~\bibnamefont{Conrad}},
  \bibinfo{journal}{Nucl. Instrum. Meth.} \textbf{\bibinfo{volume}{A551}},
  \bibinfo{pages}{493} (\bibinfo{year}{2005}), \arxiv{physics/0403059}.

\bibitem[{\citenamefont{Cowan et~al.}(2010)\citenamefont{Cowan, Cranmer, Gross,
  and Vitells}}]{Cowan:2010js}
\bibinfo{author}{\bibfnamefont{G.}~\bibnamefont{Cowan}},
  \bibinfo{author}{\bibfnamefont{K.}~\bibnamefont{Cranmer}},
  \bibinfo{author}{\bibfnamefont{E.}~\bibnamefont{Gross}}, \bibnamefont{and}
  \bibinfo{author}{\bibfnamefont{O.}~\bibnamefont{Vitells}}
  (\bibinfo{year}{2010}), \arxiv{1007.1727}.

\bibitem[{\citenamefont{Junk}(1999)}]{Junk:1999kv}
\bibinfo{author}{\bibfnamefont{T.}~\bibnamefont{Junk}}, \bibinfo{journal}{Nucl.
  Instrum. Meth.} \textbf{\bibinfo{volume}{A434}}, \bibinfo{pages}{435}
  (\bibinfo{year}{1999}), \arxiv{hep-ex/9902006}.

\bibitem[{\citenamefont{Read}(2002)}]{Read:2002hq}
\bibinfo{author}{\bibfnamefont{A.~L.} \bibnamefont{Read}}, \bibinfo{journal}{J.
  Phys.} \textbf{\bibinfo{volume}{G28}}, \bibinfo{pages}{2693}
  (\bibinfo{year}{2002}).

\bibitem[{\citenamefont{Wilks}(1938)}]{Wilks:1938ll}
\bibinfo{author}{\bibfnamefont{S.~S.} \bibnamefont{Wilks}},
  \bibinfo{journal}{Ann. Math. Statist.} \textbf{\bibinfo{volume}{9}},
  \bibinfo{pages}{60} (\bibinfo{year}{1938}).

\bibitem[{\citenamefont{Arneodo et~al.}(2000)}]{Arneodo:2000vc}
\bibinfo{author}{\bibfnamefont{F.}~\bibnamefont{Arneodo}} \bibnamefont{et~al.},
  \bibinfo{journal}{Nucl. Instrum. Meth.} \textbf{\bibinfo{volume}{A449}},
  \bibinfo{pages}{147} (\bibinfo{year}{2000}).

\bibitem[{\citenamefont{Akimov et~al.}(2002)}]{Akimov:2001pb}
\bibinfo{author}{\bibfnamefont{D.}~\bibnamefont{Akimov}} \bibnamefont{et~al.},
  \bibinfo{journal}{Phys. Lett.} \textbf{\bibinfo{volume}{B524}},
  \bibinfo{pages}{245} (\bibinfo{year}{2002}), \arxiv{hep-ex/0106042}.

\bibitem[{\citenamefont{Bernabei et~al.}(2001)}]{Bernabei:2001pz}
\bibinfo{author}{\bibfnamefont{R.}~\bibnamefont{Bernabei}}
  \bibnamefont{et~al.}, \bibinfo{journal}{Eur. Phys. J. direct}
  \textbf{\bibinfo{volume}{C3}}, \bibinfo{pages}{11} (\bibinfo{year}{2001}).

\bibitem[{\citenamefont{Aprile et~al.}(2005)}]{Aprile:2005mt}
\bibinfo{author}{\bibfnamefont{E.}~\bibnamefont{Aprile}} \bibnamefont{et~al.},
  \bibinfo{journal}{Phys. Rev.} \textbf{\bibinfo{volume}{D72}},
  \bibinfo{pages}{072006} (\bibinfo{year}{2005}), \arxiv{astro-ph/0503621}.

\bibitem[{\citenamefont{Chepel et~al.}(2006)}]{Chepel:2006yv}
\bibinfo{author}{\bibfnamefont{V.}~\bibnamefont{Chepel}} \bibnamefont{et~al.},
  \bibinfo{journal}{Astropart. Phys.} \textbf{\bibinfo{volume}{26}},
  \bibinfo{pages}{58} (\bibinfo{year}{2006}).

\bibitem[{\citenamefont{Aprile et~al.}(2009)}]{Aprile:2008rc}
\bibinfo{author}{\bibfnamefont{E.}~\bibnamefont{Aprile}} \bibnamefont{et~al.},
  \bibinfo{journal}{Phys. Rev.} \textbf{\bibinfo{volume}{C79}},
  \bibinfo{pages}{045807} (\bibinfo{year}{2009}), \arxiv{0810.0274}.

\bibitem[{\citenamefont{Manzur et~al.}(2010)}]{Manzur:2009hp}
\bibinfo{author}{\bibfnamefont{A.}~\bibnamefont{Manzur}} \bibnamefont{et~al.},
  \bibinfo{journal}{Phys. Rev.} \textbf{\bibinfo{volume}{C81}},
  \bibinfo{pages}{025808} (\bibinfo{year}{2010}), \arxiv{0909.1063}.

\bibitem[{\citenamefont{Plante et~al.}(2011)}]{Plante:2011hw}
\bibinfo{author}{\bibfnamefont{G.}~\bibnamefont{Plante}} \bibnamefont{et~al.}
  (\bibinfo{year}{2011}), \arxiv{1104.2587}.

\bibitem[{\citenamefont{Bezrukov et~al.}(2010)\citenamefont{Bezrukov,
  Kahlhoefer, and Lindner}}]{Bezrukov:2010qa}
\bibinfo{author}{\bibfnamefont{F.}~\bibnamefont{Bezrukov}},
  \bibinfo{author}{\bibfnamefont{F.}~\bibnamefont{Kahlhoefer}},
  \bibnamefont{and} \bibinfo{author}{\bibfnamefont{M.}~\bibnamefont{Lindner}}
  (\bibinfo{year}{2010}), \arxiv{1011.3990}.

\bibitem[{\citenamefont{Szydagis et~al.}(2011)}]{Szydagis:2011tk}
\bibinfo{author}{\bibfnamefont{M.}~\bibnamefont{Szydagis}} \bibnamefont{et~al.}
  (\bibinfo{year}{2011}), \arxiv{1106.1613}.

\bibitem[{\citenamefont{Lewin and Smith}(1996)}]{Lewin:1995rx}
\bibinfo{author}{\bibfnamefont{J.~D.} \bibnamefont{Lewin}} \bibnamefont{and}
  \bibinfo{author}{\bibfnamefont{P.~F.} \bibnamefont{Smith}},
  \bibinfo{journal}{Astropart. Phys.} \textbf{\bibinfo{volume}{6}},
  \bibinfo{pages}{87} (\bibinfo{year}{1996}), \bibinfo{note}{{see also
  \textit{Detector response corrections: correction} and \textit{Spin factors -
  revised tables}, available from
  \texttt{http://hepwww.rl.ac.uk/UKDMC/pub/publications.html}}}.

\bibitem[{\citenamefont{Smith et~al.}(2007)}]{Smith:2006ym}
\bibinfo{author}{\bibfnamefont{M.~C.} \bibnamefont{Smith}}
  \bibnamefont{et~al.}, \bibinfo{journal}{Mon. Not. Roy. Astron. Soc.}
  \textbf{\bibinfo{volume}{379}}, \bibinfo{pages}{755} (\bibinfo{year}{2007}),
  \arxiv{astro-ph/0611671}.

\bibitem[{\citenamefont{Aprile et~al.}(2011)}]{Aprile:2011vb}
\bibinfo{author}{\bibfnamefont{E.}~\bibnamefont{Aprile}} \bibnamefont{et~al.},
  \bibinfo{journal}{Phys. Rev.} \textbf{\bibinfo{volume}{D83}},
  \bibinfo{pages}{082001} (\bibinfo{year}{2011}), \arxiv{1101.3866}.

\bibitem[{\citenamefont{Trotta et~al.}(2008)\citenamefont{Trotta, Feroz,
  Hobson, Roszkowski, and Ruiz~de Austri}}]{Trotta:2008bp}
\bibinfo{author}{\bibfnamefont{R.}~\bibnamefont{Trotta}},
  \bibinfo{author}{\bibfnamefont{F.}~\bibnamefont{Feroz}},
  \bibinfo{author}{\bibfnamefont{M.~P.} \bibnamefont{Hobson}},
  \bibinfo{author}{\bibfnamefont{L.}~\bibnamefont{Roszkowski}},
  \bibnamefont{and} \bibinfo{author}{\bibfnamefont{R.}~\bibnamefont{Ruiz~de
  Austri}}, \bibinfo{journal}{JHEP} \textbf{\bibinfo{volume}{12}},
  \bibinfo{pages}{024} (\bibinfo{year}{2008}), \arxiv{0809.3792}.

\bibitem[{\citenamefont{Aalseth et~al.}(2010)}]{Aalseth:2010vx}
\bibinfo{author}{\bibfnamefont{C.~E.} \bibnamefont{Aalseth}}
  \bibnamefont{et~al.} (\bibinfo{year}{2010}), \arxiv{1002.4703}.

\bibitem[{\citenamefont{Savage et~al.}(2009)\citenamefont{Savage, Gelmini,
  Gondolo, and Freese}}]{Savage:2008er}
\bibinfo{author}{\bibfnamefont{C.}~\bibnamefont{Savage}},
  \bibinfo{author}{\bibfnamefont{G.}~\bibnamefont{Gelmini}},
  \bibinfo{author}{\bibfnamefont{P.}~\bibnamefont{Gondolo}}, \bibnamefont{and}
  \bibinfo{author}{\bibfnamefont{K.}~\bibnamefont{Freese}},
  \bibinfo{journal}{JCAP} \textbf{\bibinfo{volume}{0904}}, \bibinfo{pages}{010}
  (\bibinfo{year}{2009}), \arxiv{0808.3607}.

\bibitem[{\citenamefont{Bernabei et~al.}(2010)}]{Bernabei:2010mq}
\bibinfo{author}{\bibfnamefont{R.}~\bibnamefont{Bernabei}}
  \bibnamefont{et~al.}, \bibinfo{journal}{Eur. Phys. J.}
  \textbf{\bibinfo{volume}{C67}}, \bibinfo{pages}{39} (\bibinfo{year}{2010}),
  \arxiv{1002.1028}.

\end{thebibliography}
%\end{document}
% manual changes from xenonprofile.bbl:
% - replaced \eprint by \arxiv
% - replaced \emph by \textit
% - reordered yellin

\end{document}